\documentclass[aps,prl,amsmath,reprint,superscriptaddress,nofootinbib]{revtex4-1}
\usepackage{graphicx}
\usepackage{hyperref}
\usepackage{color}
\usepackage{ulem}

\renewcommand{\Im}{\operatorname{Im}}

\newcommand{\epsPar}{\epsilon_{\parallel}}
\newcommand{\epsPerp}{\epsilon_{\perp}}
\newcommand{\citeasnoun}[1]{Ref.~\onlinecite{#1}}
\renewcommand{\eqref}[1]{Eq.~(\ref{eq:#1})}
\newcommand{\eqreftwo}[2]{Eqs.~(\ref{eq:#1},\ref{eq:#2})}

\newcommand{\added}[1] {#1}
\newcommand{\removed}[1] {}

\begin{document}

\normalem 

\title{The effectiveness of thin films in lieu of hyperbolic metamaterials in the near field}

\author{Owen D. Miller}
\affiliation{Department of Mathematics, Massachusetts Institute of
  Technology, Cambridge, MA 02139}
\author{Steven G. Johnson}
\affiliation{Department of Mathematics, Massachusetts Institute of
  Technology, Cambridge, MA 02139}
\author{Alejandro W. Rodriguez}
\affiliation{Department of Electrical Engineering, Princeton
  University, Princeton, NJ 08544}

\begin{abstract}
  We show that the near-field functionality of hyperbolic
  metamaterials (HMM), typically proposed for increasing the photonic local density of states (LDOS), can be achieved with thin metal films.  \removed{Essentially, relative to thin films, HMMs have a broader but weaker coupling to the high-wavevector states of near-field emitters, such that when integrated over wavevector the overall coupling of the two is essentially equivalent. }\added{Although HMMs have an infinite density of internally-propagating plane-wave states, the external coupling to nearby emitters is severely restricted.}
We show analytically that properly designed thin films, of thicknesses comparable to the metal size of a hyperbolic metamaterial, yield an LDOS as high as (if not higher than) that of HMMs. We illustrate these ideas by performing exact numerical computations of the LDOS of multilayer HMMs, along with their application to the problem of maximizing near-field heat transfer, to show that \added{single-layer }thin films are suitable replacements in both cases.
\end{abstract}

\maketitle

Near-field optics involves coupling of evanescent waves and holds
great promise for applications ranging from fluorescent imaging
\cite{Betzig1992,VanZanten2009,Schermelleh2010} to thermophotovoltaic
heat transfer \cite{Laroche2006,Basu2009}. Evanescent waves from
nearby radiative emitters can couple, for example, to plasmon modes at
metal-dielectric interfaces \cite{Morawitz1974}, surface states in
photonic crystals \cite{Luo2003}, and, as recently proposed, to a
continuum of propagating modes in effective, anisotropic materials
with hyperbolic dispersion~\cite{Krishnamoorthy2012,Guo2013}.  In this
letter, we show that the local density of states (LDOS) near a hyperbolic
metamaterial (HMM) \cite{Krishnamoorthy2012,Elser2006,Biehs2012,Cortes2012,Guo2012a}, even in the perfect effective--medium limit, is
fundamentally no larger than the LDOS near thin metal films.  Despite
the HMM states being accessible for almost all high--wavevector waves,
their coupling strength to near-field electric dipoles is only
moderate.  We show analytically that thin metal films, whose resonant
waves couple very strongly at a small number of large-wavevector
states, yield an equally large LDOS upon integration over wavevector.
Moreover, the film thickness required to match the operational
frequency of the HMM is of the same order of magnitude as the size of
the metal within the HMM, such that the thin film is much easier to fabricate.  Although we begin with an idealized asymptotic analysis to illustrate the basic physics, we confirm these conclusions with exact calculations of LDOS and heat transfer in realistic materials, obtaining comparable results for HMMs and thin films.

Hyperbolic metamaterials are periodic, metallodielectric composites with a unit cell size $a$ much smaller than the wavelength, simplifying their electromagnetic response to that of a homogeneous medium \cite{Krishnamoorthy2012,Elser2006,Poddubny2013}.  Typical structures have \added{one- or two-dimensional }periodicity\removed{ in only one or two dimensions}, yielding an anisotropic effective--permittivity tensor that, for certain materials and dimensions, has components $\epsPar$ (surface--parallel) and $\epsPerp$ with opposite signs ($\epsPar\epsPerp<0$).  Such a material has hyperbolic dispersion, leading to propagating plane wave \removed{solutions}\added{modes} with parallel wavevector larger than the free-space wavevector, i.e. $k_\Vert > \omega/c$, for frequency $\omega$.  They have excited great interest because of their potential to increase the local density of states (LDOS)~\cite{Jacob2010,Noginov2010,Kidwai2011,Ni2011,Jacob2012,Kim2012,Guo2012,Narimanov2012, Biehs2012,Guo2012a,Cortes2012,Guo2013,Liu2013,Poddubny2013,Lu2014} , e.g. for radiative--lifetime engineering \cite{Jacob2010,Noginov2010,Kidwai2011,Ni2011,Jacob2012,Kim2012} and near-field
heat transfer enhancement
\cite{Guo2012,Biehs2012,Narimanov2012,Guo2013,Liu2013}.

\added{Previous works have explained the increase in the LDOS as arising from the hyperbolic $\omega$-$k$ dispersion relation, which implies an infinite density of states (DOS) at all frequencies for which $\epsPar\epsPerp<0$.  Typically \cite{comment}, the proposed metamaterials exhibit orders of magnitude enhancements for the LDOS, or for radiative heat transfer, when compared to vacuum \cite{Jacob2012,Guo2012a}, bulk metal \cite{Jacob2010,Noginov2010,Kidwai2011,Ni2011,Biehs2012,Liu2013}, or blackbody \cite{Guo2012,Biehs2012,Narimanov2012,Guo2013,Liu2013} systems.  We will show, however, that the increased LDOS in each case is not due to anisotropy, but rather to the reduction in resonant frequency that arises when the fraction of metal is reduced.  As we show below, an effective metamaterial with \emph{isotropic} $\epsilon \approx -1$, which achieves the resonance shift without anisotropy, is better than the ideal HMM with oppositely signed permittivity components.  While such an isotropic metamaterial would likely require complex three-dimensional fabrication \cite{Pendry1996,Sievenpiper1996}, here we show that thin films, well-studied systems with resonance frequencies far below the bulk plasma frequency $\omega_p$ \cite{Economou1969,Kliewer1967,Ritchie1973,Burke1986,Francoeur2008,
Biehs2007,Ben-Abdallah2009,Basu2011}, exhibit the same near-field functionality as HMMs.  }\removed{Typically, the LDOS is compared to vacuum or to bulk materials, and
the heat transfer is compared to far-field blackbodies or to bulk
metals. Yet it is well understood that vacuum and/or blackbodies are not ideal in the near field,
and neither are bulk metals when operating far from the plasma
frequency $\omega_p$ of the underlying metal.  A better comparison is
to thin metal films, which \emph{can} exhibit resonances at
frequencies much smaller than $\omega_p$.  Thin films are very
well studied and well understood, and when
perforated lead to extraordinary transmission.  Here we directly compare the
two structures, hyperbolic metamaterials on the one hand and thin
films on the other, and show that the functionally of these two
systems can be very similar in the near field.  }A primary difference
is the larger bandwidth provided by thin films; conversely, one could
say that HMMs offer selectivity.  \removed{It should be noted, h}\added{H}owever, \removed{that
}selectivity is a general property of
metamaterials~\cite{Pendry1996,Sievenpiper1996,Sievenpiper1999,Petersen2013} \removed{or any material based on resonant particles }and
does not arise from hyperbolicity.

As a means for comparing the HMM to the thin film, we encapsulate the
response of either structure in a scattering matrix $S(k_\Vert,\omega)$. Such a
description is valid for any linear system with translational and rotational
symmetry, including a media stack with many layers, an effective
material with anisotropic permittivity, or a single thin film.
Although the ultimate quantity of interest is the LDOS
$\rho(z,\omega)$ at a point $z$ near the interface, as in
\citeasnoun{Guo2012} we will define the weighted local density of
states (WLDOS) $\rho(z,\omega,k_\Vert)$ by $\rho(z,\omega) = \int
\rho(z,\omega,k_\Vert)\,dk_\Vert$, thereby resolving the contribution at each
$k_\Vert$.  In the near field ($k_\Vert \sim 1/z \gg \omega/c$), the WLDOS of an electric dipole is
dominated by contributions from the p (TM) polarization, given by \cite{Wijnands1997,Joulain2003}:
\begin{align}
\rho(k_\Vert,\omega,z) \approx \frac{1}{2\pi^2 \omega} k_\Vert^2 e^{-2 k_\Vert z} \Im S_{21}\left(k_\Vert,\omega\right) .
\label{eq:WLDOS}
\end{align}
\removed{Hence, r}\added{R}egardless of the origin of the large-wavevector states,
i.e. from either a continuum of propagating hyperbolic modes or a
discrete set of plasmonic modes, the key to increasing the LDOS is to
increase $\Im S_{21}$\added{, the imaginary part of the reflection coefficient.  Thus, even if the DOS \emph{within} a structure is infinite, as in an ideal HMM, the local density of states (LDOS) near the structure additionally requires strong external coupling.  We will see that HMMs have only moderate external coupling, with $\Im S_{21} \leq 1$, limiting their total LDOS.}

\emph{\added{Anisotropic permittivity and hyperbolic dispersion}\removed{Ideal metal vs. ideal HMM}}: \added{To isolate the contribution of anisotropy, without any shift in resonance, we first compare the ideal anisotropic material, with hyperbolic dispersion, to the ideal isotropic metallic permittivity, which supports surface plasmon modes.}\removed{We begin with a motivating example
to understand whether hyperbolic modes, propagating within the
metamaterial and evanescent to the exterior, are fundamentally
superior to plasmon modes bound to the surface of a metal.  }  We assume
a single interface, with vacuum on one side and a bulk material on the
other.  Forgoing fabrication concerns for the moment, we ask what
material provides the largest near-field LDOS at large parallel
wavevector $k_\Vert \gg \omega/c$?  \removed{It is well-known that a}\added{A} surface
plasmon \removed{occurs }at a metal--vacuum interface \removed{if }\added{exhibits maximum DOS for }$\epsilon_{\textrm{metal}}=-1$ \cite{Maier}.  Similarly, the \removed{optimal anisotropic
permittivity}\added{largest LDOS occurs} for an HMM \removed{has}\added{with} $\epsPar = -1$ and $\epsPerp = 1$ \cite{Guo2012,Guo2012a} (or
vice versa, at large $k_\Vert$ only the product
matters \added{and there is no distinction between Type I and Type II HMMs}).  We can add any amount of loss $\epsilon_i$ to the
permittivities, \removed{such that}\added{defining} the permittivities \removed{are}\added{to be}:
\begin{align}
\label{eq:eps_metal}
\epsilon_{\textrm{ideal metal}} &= -1 + i\epsilon_i \\
\epsilon_{\textrm{ideal HMM}} & = 
\begin{pmatrix} 
-1 & 0 & 0 \\
0 & -1 & 0 \\
0 & 0 & 1 
\end{pmatrix} + i\epsilon_i
\label{eq:eps_hmm}
\end{align}
\removed{We can write t}\added{T}he imaginary part\added{s}  of the (TM) reflectivity $S_{21}$ for the ideal metal and HMM \removed{as}\added{are}
\begin{align}
\Im\left(S_{21}\right)_{\textrm{ideal metal}} &= \Im\left( \frac{\epsilon - 1}{\epsilon + 1}\right) = \frac{2}{\epsilon_i} \\
\Im\left(S_{21}\right)_{\textrm{ideal HMM}} &= \Im\left( \frac{\epsPar\epsPerp - \sqrt{\epsPar\epsPerp}}{\epsPar\epsPerp + \sqrt{\epsPar\epsPerp}} \right) \\
&= \frac{2 \sqrt{1 + \epsilon_i^2}}{2 + \epsilon_i^2}
\end{align}
where the reflectivities are independent of $k_\Vert$ in the limit $k_\Vert \gg \omega/c$.  The ratio of the respective LDOS is then:
\begin{align}
\frac{\rho_{\textrm{ideal metal}}(k_\Vert,\omega)}{\rho_{\textrm{ideal HMM}}(k_\Vert,\omega)} = \sqrt{1 + 1/\epsilon_i^2} + \frac{1}{\epsilon_i \sqrt{1 + \epsilon_i^2}} > 1
\end{align}
where we see that regardless of the loss value, the ideal metal, with
$\epsilon = -1 + i\epsilon_i$, is always better than the HMM with $\epsilon =
(\epsPar,\epsPerp) = (-1,1) + i\epsilon_i$.  \added{There should be no difference in photon lifetimes, as in each case the complex wavevector of either the surface plasmon or HMM mode is of the form $k_\Vert \approx \left[\left(1 + i\right) / \sqrt{2} \epsilon_i\right]\omega / c$.  }We note that the choice
of permittivity in \eqreftwo{eps_metal}{eps_hmm} is exactly optimal
only for $\epsilon_i = 0$, but that is true for both structures, and
thus their relative performance is still meaningful.  Moreover, in the
exactly optimal limit as $\epsilon_i \rightarrow 0$, the metal's LDOS
diverges, whereas the HMMs remains finite.

\emph{\added{HMM vs. thin film}\removed{Reflectivity--bandwidth product}}: Away from the surface--plasmon
frequency where $\epsilon \approx -1$, can metals still compete with
HMMs?  Bulk metals cannot, since their coupling to the
single--interface surface plasmon is weak.  Alternatively, it is
well known that a thin metallic slab couples the front- and
rear-surface plasmons \cite{Maier,Economou1969,Kliewer1967,Ritchie1973}, yielding a symmetric mode that can exist at
$\omega \ll \omega_p$ even for large $k_\Vert$.  The thin film modes still
asymptotically approach $\omega_p/\sqrt{2}$ as $k_\Vert \rightarrow \infty$, but if $k_{\textrm{max}}$ is the maximum $k_\Vert$ of interest---defined in HMMs by the unit
cell---a film can exhibit low--frequency modes at $k_\Vert \sim k_{\textrm{max}}$.

\begin{figure}
\includegraphics[width=\linewidth]{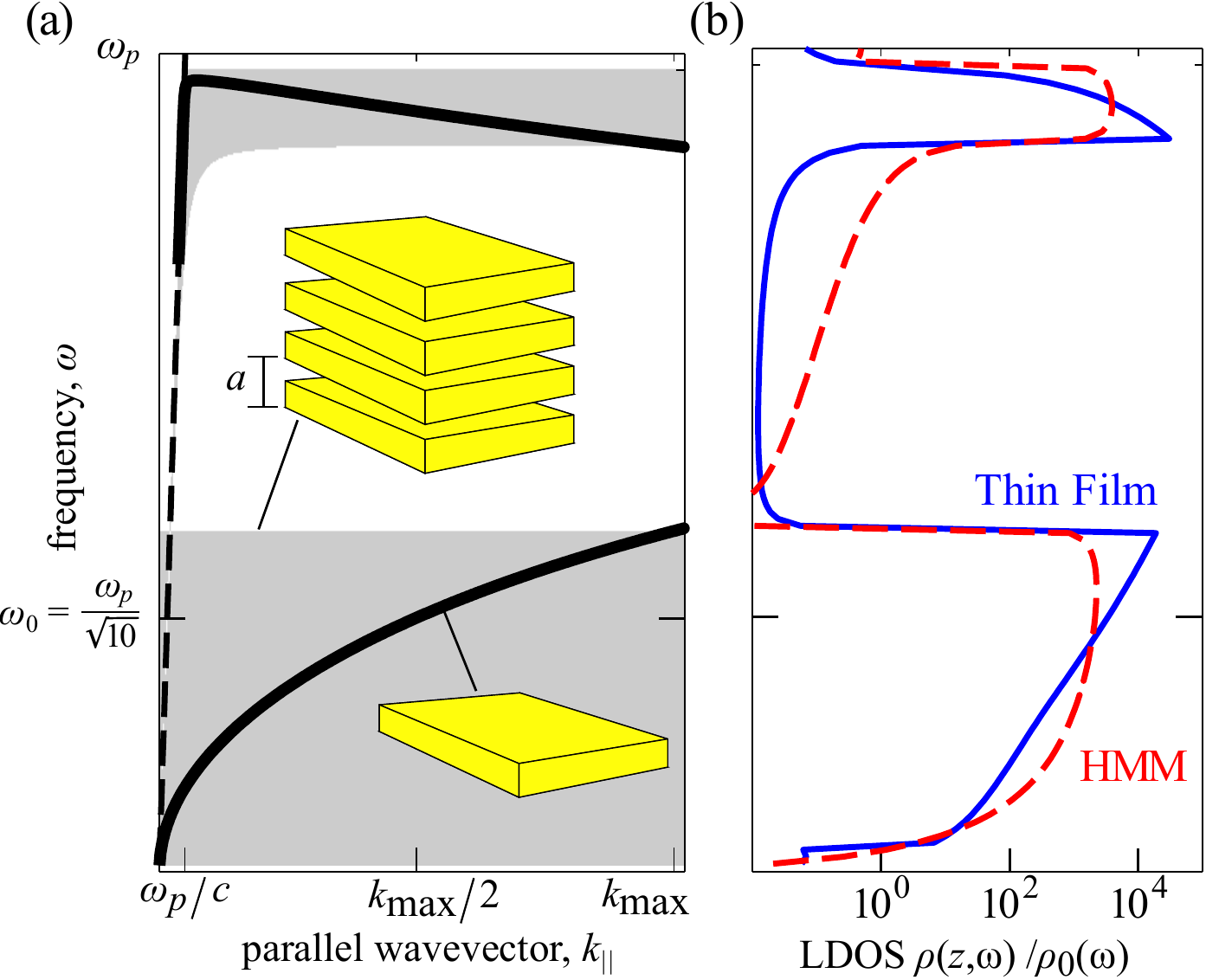}
\caption{\label{fig1}(a) Comparison of the continuum of modes in an HMM to
  the discrete modes of a thin film.  The surface modes of thin films
  have two branches, corresponding to symmetric and antisymmetric
  modes, that split away from $\omega_p/\sqrt{2}$.  The HMM is
  designed for maximum LDOS at $\omega_0 = \omega_p/\sqrt{10}$, while
  the film is designed to have a resonance at
  $(\omega_0,k_{\textrm{max}}/2)$.  The HMM comprises
  alternating layers of a lossless Drude metal (plasma frequency
  $\omega_p$) and dielectric $\epsilon=1$, with metallic fill fraction
  $f$ given by \eqref{FF_HMM}.  The thin film consists of the same
  metal, with thickness $d$ given by \eqref{D_TF}.  The unit cell $a = 0.1c/\omega_p$ ($k_{\textrm{max}}c/\omega_p=20\pi$).  HMMs exhibit
  $\Im\left(S_{21}\right) \approx 1$ (shading indicates
  $\Im\left(S_{21}\right)>0$) for many $k_\Vert$, whereas thin films
  provide resonances with $\Im\left(S_{21}\right) \gg 1$ for smaller
  bandwidths $\Delta k_\Vert$.  (b) LDOS for each structure (normalized to
  the vacuum LDOS $\rho_0$) at $z=a$, the closest point at which EMT is valid.  Note that
  the two structures have almost equal LDOS at $\omega_0$, as predicted
  by \eqreftwo{Prod_HMM}{Prod_TF}.}
\end{figure}

For the thin metal film to be a practical replacement for hyperbolic
metamaterials, the optimal structure must not be too thin.  We now
analyze a second case: optimizing the near-field LDOS over a band of
frequencies centered at $\omega_0$, for a lossless metal with
permittivity $\epsilon(\omega_0) = \epsilon_m \ll -1$ (to contrast
with the $\epsilon=-1$ case studied previously).  We will design an
optimal HMM and an optimal metallic thin film, and find that the LDOS
of each is roughly equal.

We consider HMMs composed of a metal with permittivity $\epsilon_m$
and a dielectric with permittivity $\epsilon=1$ (for simplicity),
with a metallic fill fraction $f$.  Typical
effective-medium theory (EMT) approximations of HMMs assume multilayer
slabs~\cite{Guo2012,Cortes2013} or periodic
cylinders~\cite{Elser2006,Biehs2012,Cortes2013}.  For either one, the optimal fill fraction is given by
\begin{align}
f \approx \frac{2}{|\epsilon_m|},
\label{eq:FF_HMM}
\end{align} 
\removed{which}\added{chosen to} satisf\added{y}\removed{ies} \added{the ideal relation} $\epsPar\cdot \epsPerp \approx -1$ at $\omega_0$.  We now
have the lossless version of the ideal scenario analyzed before, yielding the imaginary part of the reflectivity as
\begin{align}
\Im \left(S_{21}\right)_{\textrm{HMM}} = 1,
\label{eq:S21_HMM}
\end{align}
in agreement with previously derived results \cite{Guo2012}.  Although
\eqref{S21_HMM} is independent of wavevector, ultimately $a$, the size
of the unit cell within the HMM, limits the EMT approximation to $k
\lesssim k_{\textrm{max}} = 2\pi / a$ (another limiting factor is
$2\pi/z$), such that the bandwidth of the contribution to the LDOS is $\Delta k_\Vert \approx 2\pi/a$.  

We can similarly derive the optimal thin film structure, comprising
the same metal with permittivity $\epsilon_m$, and thickness $d$.  To
roughly match the performance of the HMM, we design the thin film to
have a mode at $\omega=\omega_0$ and $k_{\textrm{res,tf}} =
k_{\textrm{max}} / 2 = \pi/a$.  In agreement with the choice of
dielectric within the HMM, we assume vacuum at the rear surface of the
film.  The reflectivity of a thin film \cite{Coldren2012} is given by
$S_{21} = r_{01}\left[1-\operatorname{exp}(-k_\Vert d)\right] / \left[1 -
  r_{01}^2\operatorname{exp}(-2k_\Vert d)\right]$, where $r_{01}$ is the
reflectivity at the air--metal interface.  It follows that the optimal
thickness is given by:
\begin{align}
d = \frac{a}{\pi} \ln |r_{01}| \approx \frac{2a}{\pi|\epsilon_m|}
\approx \frac{af}{\pi},
\label{eq:D_TF}
\end{align}
which represents a pole in the reflectivity spectrum.

Hence, we see that the optimal thickness is within a factor $\pi$ of
$af$; that is, it scales with the size of the metal in the HMM.  In a
multilayer HMM $af$ is exactly the thickness of the individual metal
layers, while in e.g. nanorod HMMs, $af$ is the individual nanorod
radius multiplied by the square root of the fill fraction.  \added{Because the thin film has approximately the thickness of the metal within a single unit cell, the thin film will have less or nearly equal absorptive losses as compared to the full HMM structure}.

\begin{figure*}
\centering \includegraphics[width=\linewidth]{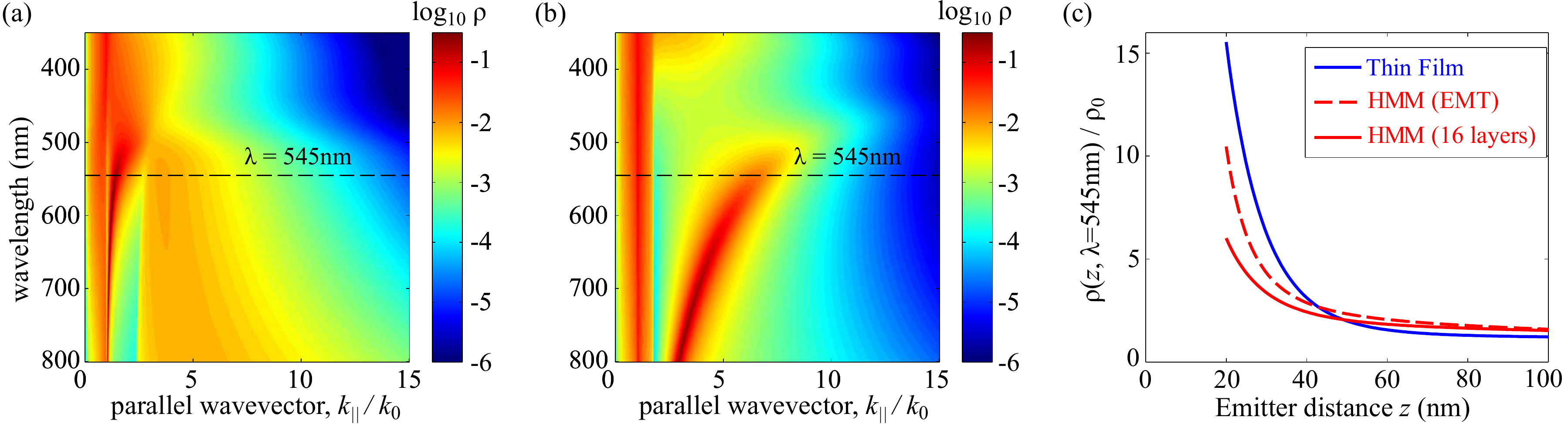}
\caption{\label{fig2}(a,b) WLDOS computations ($z=30$nm) for (a) Ag/AlO$_2$ HMM
  described by EMT and (b) Ag thin film on AlO$_2$ substrate, illustrating the
  distinct contrast between a continuum of propagating modes in the
  HMM and a single plasmonic mode in the film.  For convenience,
  $\rho$ is multiplied by $ac$, where $c$ is the speed of light.  (c)
  LDOS (normalized to the vacuum LDOS $\rho_0$) at $\lambda=545$nm for
  the thin film, the HMM (EMT), and a 16-layer implementation
  of the HMM ($a=30$nm, $f=0.4$).}
\end{figure*}

To compute the bandwidth for the thin film, we
add a loss $\epsilon''$ to the permittivity to avoid singular poles in
the reflectivity, and then take the limit $\epsilon'' \rightarrow 0$.
Since $k_\Vert d \ll 1$ (which follows from $\epsilon_m \ll -1$), we can
approximate $\operatorname{exp}(-2k_\Vert d) \approx 1 - 2k_\Vert d$.  On
resonance, the imaginary part of the reflectivity is given by
\begin{align}
\Im \left(S_{21}\right)_{\textrm{thin film}} =\frac{2}{\epsilon''k_{\textrm{res,tf}}d} = \frac{|\epsilon_m|}{\epsilon''}
\label{eq:S21_TF}
\end{align}
The full-width half-max bandwidth $\Delta k_\Vert$ of $\Im S_{21}$ is
\begin{align}
\Delta k_\Vert \approx \frac{2\pi}{a}\frac{\epsilon''}{|\epsilon_m|}.
\label{eq:deltaK_TF}
\end{align}
The LDOS at $\omega_0$ requires a full integration of \eqref{WLDOS}, but we can define a simpler ``reflectivity--bandwidth''
product to approximate the contribution of the reflectivity to the integral (verifying later the accuracy of the approximation).  From Eqs.~(\ref{eq:S21_HMM},\ref{eq:S21_TF},\ref{eq:deltaK_TF}), valid in the limits $k_\Vert \gg \omega/c$ and $|\epsilon_m| \gg 1$, we have:
\begin{align}
\left[ \Im\left(S_{21}\right)\Delta k_\Vert \right]&_{\textrm{HMM}} \approx k_{\textrm{max}} = \frac{2\pi}{a}
\label{eq:Prod_HMM} \\
\left[ \Im\left(S_{21}\right)\Delta k_\Vert \right]&_{\textrm{thin film}} \approx \frac{2\pi}{a}.
\label{eq:Prod_TF}
\end{align}
Thus, given an optimal HMM, a thin film can designed without
further fabrication difficulty and with approximately equal increase
in LDOS.

Figure~\ref{fig1} clarifies the similarities between HMMs and metallic
thin films.  A lossless Drude metal is employed for both an optimal HMM
and an optimal thin film. The center frequency is $\omega_0 = \omega_p /
\sqrt{10}$, and the unit cell $a$ is chosen to be $a=0.1c/\omega_p$.  The HMM fill fraction and thin film thickness are
chosen according to \eqreftwo{FF_HMM}{D_TF}.  The multilayer EMT is
used (a nanorod model only shifts the upper band of states downward).
The computations are exact and do not include any of the high-$k$
approximations utilized in the analysis.  Although the underlying
modes are very different---a continuum of propagating modes in HMMs
versus discrete guided modes for thin films---their \removed{overall
contribution to the }LDOSs near $\omega_0$ \removed{is }\added{are approximately }equal\added{.}\removed{, up to constants of
order one.}

\emph{Comparisons of LDOS and heat transfer}: We now move from
asymptotic analytical results, which reveal the underlying
physics, to rigorous computations of the LDOS and near-field heat
transfer characteristics for real material systems.  We assume
multilayer implementations of the HMMs, which enables us to use the
exact Green's functions \cite{Sipe1987} in both
computations~\cite{codes_github}, using the numerically stable
scattering--matrix formalism \cite{Francoeur2009}.

Figure~\ref{fig2} compares the near-field LDOS for an HMM comprising
Ag/AlO$_2$ (similar to \citeasnoun{Cortes2013}), with that of a thin
silver film on an AlO$_2$ substrate.  In the WLDOS plots, one observes
the contrast between the relatively large number of weakly--coupled
HMM modes and the single, strongly-coupled plasmonic mode of the film.
The integrated LDOS at $\lambda=545$nm is shown in Fig.~\ref{fig2}(c),
which also includes a 16-layer implementation of the HMM, with a unit
cell of $30$nm and fill fraction $f=0.4$ (chosen to approximately
maximize the contribution of hyperbolic modes).  The unit cell defines
the minimum $z$ at which EMT is applicable, no smaller than $z=20$nm.  One can see that the thin film (thickness $=8$nm) has
a larger LDOS in the near field.

\begin{figure}
\includegraphics[width=0.95\columnwidth]{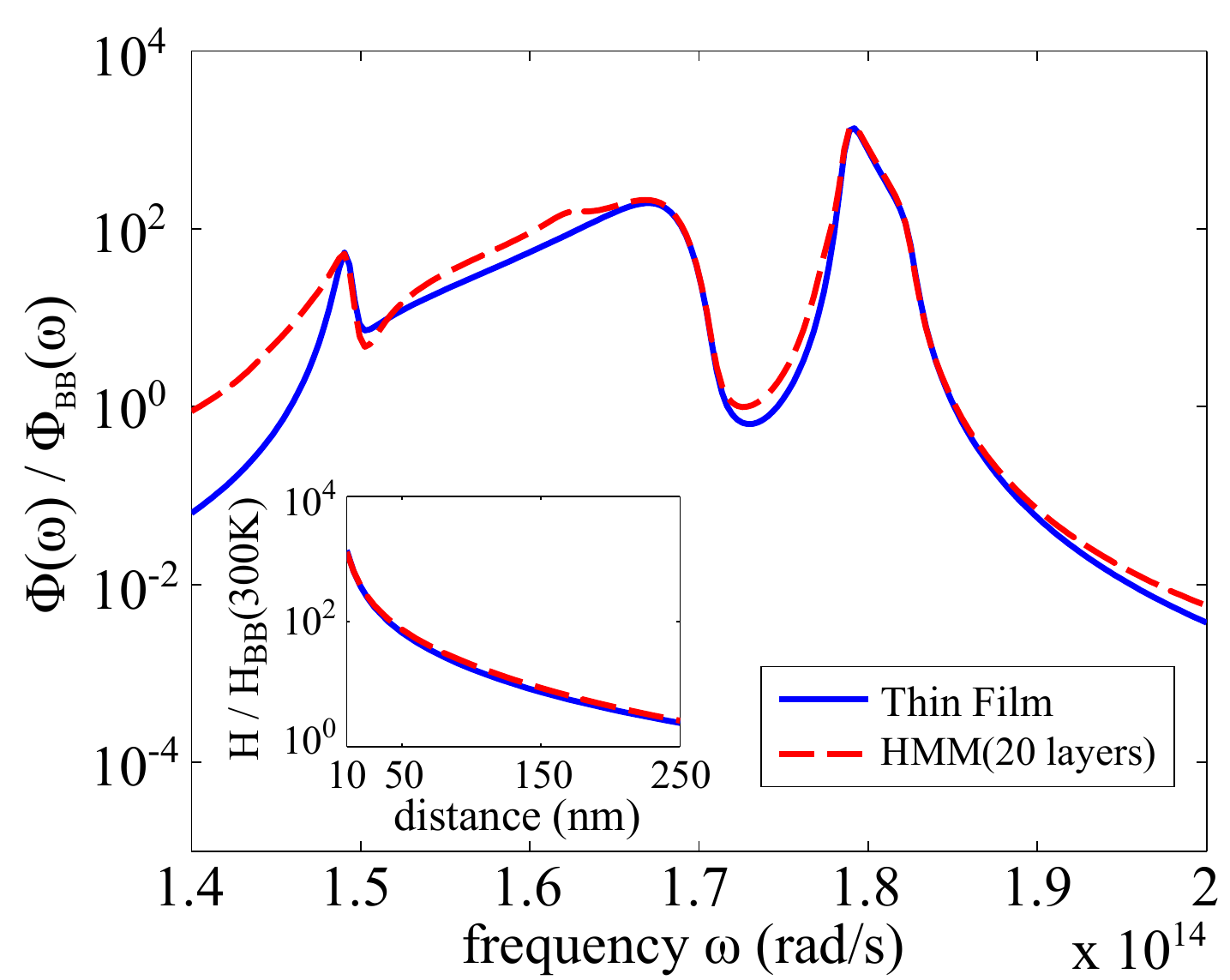}
\caption{\label{fig3} Near-field heat transfer between SiC/SiO$_2$
  structures.  Heat flux spectrum for the HMM (red), comprising 20
  layers (10 unit cells) with $a=200$nm and $f=0.25$. Removing all but
  the top layer yields a SiC thin film of thickness $50$nm on an
  SiO$_2$ substrate (blue). Each structure interacts with its mirror
  image, at a separation distance of $100$nm. Inset: The total heat
  transfer, with one object at $T=300$K and the other at $T=0$K.}
\end{figure}

Figure~\ref{fig3} compares the near-field heat transfer for a very
different, but commonly proposed \cite{Biehs2012,Guo2012,Liu2013,Guo2013,Biehs2013,Ben-Abdallah2009,Francoeur2010}, material system: SiC/SiO$_2$.  SiC is a phonon-polaritonic
metal with negative permittivities for $\omega \in [1.5,1.8]\times
10^{14}$Hz ($\lambda \approx 11$-$12\mu$m), which is promising for
heat--transfer applications where the peak of $300$K radiation is
$\lambda\approx 7.6\mu$m.  For the HMM we choose a 20 layer (10 unit
cell) implementation with each $200$nm unit cell consisting of $50$nm
of SiC and $150$nm of SiO$_2$ ($\epsilon=3.9$), consistent with previous work
\cite{Guo2013,Biehs2013}.  The total heat transfer between objects at
$T_1$ and $T_2$ is given by $H = \int_0^{\infty} \mathrm{d}\omega
\left[\Theta(\omega,T_1) - \Theta(\omega,T_2)\right] \Phi(\omega)$,
where $\Phi$ is the temperature-independent flux spectrum and $\Theta$
is the mean energy per oscillator \cite{Joulain2005}.  For comparison
with the HMM, we also consider a thin film system.  Instead of
optimizing the thickness, we choose $d=50$nm, such that the film is
equivalent to removing 19 intermediate layers from the HMM, leaving
only the top layer and the SiO$_2$ substrate.  Each computation solves
for the flux rate between an object and its mirror image.  We see in
Fig.~\ref{fig3} that the flux spectra for the HMM and the thin film
are nearly identical at $100$nm separation distance.  The inset, the
total heat transfer at $T_1=300$K and $T_2=0$K, shows even greater
similarity between the two.  These computations show that not only is
the thin film a suitable replacement for the HMM, but that the top
layer of the HMM is primarily responsible for the heat transfer in the
first place.  A similar effect was observed in \citeasnoun{Biehs2013},
albeit by labeling contributions within a structure rather than
comparing two different ones. We arrive at a different conclusion than \citeasnoun{Biehs2013}: rather than removing the top layer, to create a
structure with less heat transfer but a greater relative contribution
from propagating modes, we suggest simply replacing the HMM with a
single thin layer, optimized for even greater heat transfer.

\emph{Conclusion}: We have shown that thin films can operate as well
or better than HMMs for increasing LDOS and heat transfer. Although
previous works \cite{Cortes2013} have differentiated the ``radiative''
transitions of emitters near HMMs with ``quenching'' near a plasmonic
metal, there is no fundamental difference between creating a photon in
a bound thin--film guided mode, versus a high-wavevector photon that is trapped (propagating) within the HMM.  For any amount of loss, the
photon will eventually be absorbed unless some other mechanism couples it to the far field, an equally difficult task for either structure.  For any
near--field application, then, we expect thin films to suffice as a
replacement for HMMs.  Away from the near field, of course, there are effects HMMs can exhibit that thin films cannot, such as
negative refraction \cite{Noginov2009}.

\end{document}